\documentstyle[12pt,aaspp]{article}  

\begin{document}
\singlespace
\title{Fossil Signatures of Ancient Accretion Events in the Halo.}
\author{Kathryn V. Johnston, Lars Hernquist\altaffilmark{1} and
Michael Bolte.}

\affil{Board of Studies in Astronomy and Astrophysics,
U. C. Santa Cruz, Santa Cruz, CA 95064}
\altaffiltext{1}{Sloan Foundation Fellow, Presidential Faculty Fellow}
\begin{abstract}

The role that minor mergers 
have played in the formation and structure of the Milky Way
is still an open question, about which there is
much debate.
We use numerical simulations to explore the evolution of debris
from a tidally disrupted satellite, 
with the aim of developing a method that can be used to identify
and quantify signatures
of accretion in a survey of halo stars.
For a Milky Way with a spherical halo, we find
that debris from minor mergers 
can remain aligned along great circles
throughout the lifetime of the Galaxy.
We exploit this result to develop the method of Great
Circle Cell Counts (GC3), which we test by applying it 
to artificially constructed halo 
distributions. 
Our results suggest that if as few as 1\% of the stars in a halo survey are
accreted from the disruption of a single subsystem smaller than the Large
Magellanic Cloud, GC3 can recover the great circle associated with this debris.
The dispersion in GC3 can also be used to detect the presence of
structure characteristic of accretion 
in distributions containing a much smaller
percentage of material accreted from any single satellite.

\end{abstract}


\section{Introduction}
Theories describing the formation 
of the Milky Way can generally be viewed as
variations on or combinations of two scenarios. Based on the kinematics
of metal-poor halo field stars, Eggen, Lynden-Bell \& Sandage (1962;
ELS) proposed a model in which the Galaxy began with the free-fall
collapse of an approximately uniform, spherical primordial fluctuation.
The globular clusters and halo field stars were formed during the
free-fall phase, with the bulk of the original Galactic material dissipating
energy in the gas phase and falling into a rotationally supported
disk. Alternatively, to account for
the lack of a metallicity gradient in the halo and the
suggestion of a large spread in the ages of outer-halo globular
clusters, Searle \& Zinn (1978; SZ) proposed a scenario in
which the Galaxy was assembled through the gradual merging of many
sub-galactic sized clouds. 

Low-mass stars have lifespans that are greater than the age of
the Galaxy and do not dissipate orbital energy. Hence, we
have an abundance of ``fossil'' information with which to explore
various formation scenarios, through halo stars and
clusters, especially their abundances and kinematics. Commonly
employed tracers include the
kinematics of field stars as a function of [Fe/H] (as in
ELS), the age distribution of
Galactic globular clusters (e.g. Vandenberg, Bolte \& Stetson, 1990), 
trends in cluster age with 
[Fe/H] and Galactocentric radius (as in SZ), the lack of
an abundance gradient with radius or height above the disk, and the
persistence of a cold, thin disk (e.g.
T\'oth \& Ostriker, 1992).
(See Larson [1990] and Majewski [1993] for comprehensive 
reviews.)

Here, we consider the feasibility of probing
``fossil'' signatures of the formation of the Galaxy 
through structure in the phase-space distribution of halo
field stars. Our approach is motivated by
numerical simulations, which 
demonstrate that distinctive features such as
tidal tails are a generic consequence of
interactions between comparable mass galaxies (e.g. Toomre \& Toomre
1972; Barnes 1988, 1992; Hernquist 1992, 1993; Hibbard \& Mihos 1995).
Similarly, streamers can be produced along the orbit of a satellite galaxy
when stars are torn from it by tidal forces from its host
(e.g. McGlynn; 1990, Piatek \& Pryor, 1995; Johnston, Spergel \& 
Hernquist, 1995) .
If such tidal debris were to maintain spatial and kinematic
coherence for the lifetime of the Galaxy, then a halo formed
through the disruption of many SZ fragments would exhibit 
streakiness in its phase-space distribution, 
unlike one originating from a smooth, monolithic collapse
which ought to be mostly featureless.

The possibility that accretion events may leave observable signatures is
supported by
observations of satellites  of the Milky Way 
which display
tantalizing evidence for ongoing tidal interactions.  As part of their
analysis leading to the discovery of the Sagittarius dwarf
galaxy (hereafter Sgr), Ibata, Gilmore \& Irwin (1994) produced an
isopleth map of the overdensity of horizontal branch stars in the
region where the dwarf was thought to lie, which shows highly
elongated contours with axis ratios $\sim$ 3:1. Grillmair et al.  (1995)
have also reported the detection of distortions in the outer
regions of globular clusters.  Both sets of observations are consistent
with morphological disturbances  produced by tidal perturbations.  

Unfortunately, the low surface density of streamers from tidally
disturbed objects
makes their detection challenging.  Nevertheless, it
is plausible that stars or satellites that appear to be
associated spatially and/or kinematically are indeed debris from tidal
interactions. From their simulations of Sgr, Johnston, Spergel
\& Hernquist (1995) predict that the tidal streamers associated with
this object may be detectable as moving groups in the halo,
and that if a similar galaxy had been  destroyed 
by the Milky Way within the
last gigayear (Gyr), its remains should still be detectable as a moving
group today.

Observational evidence suggests that there is considerable
structure in the phase-space distribution of dwarf galaxy companions
in the halo. Lynden-Bell (1976, 1982) was
the among the first to note that most of the Milky Way's satellite
galaxies lie near two great circles passing close to the Galactic
Poles. Three dwarf spheroidal galaxies (Draco, Ursa Minor and Carina)
are in the vicinity of the great circle defined by the Small and Large
Magellanic Clouds and the Magellanic Stream (the ``Magellanic Plane''),
while another five (Fornax, Leo I, Leo II, Sculptor and Sextans) are
distinctly aligned along the ``Fornax-Leo-Sculptor Stream''.  
More recently, Lynden-Bell \& Lynden-Bell (1995) developed 
a method to systematically
search for coincidences of clusters along great circles, and 
recovered these and several other possible associations of halo
objects. Moreover, Lin \& Richer (1992), Majewski (1994)  and
Fusi-Pecci, Bellazzini, Cacciari \& Ferraro (1995) have shown that
several young halo globular clusters are not far from each of these
planes, which further supports an interesting physical
explanation for these alignments.

Substructure has also been found in the stellar distribution in the halo.
Doinidis \& Beers (1989) calculated the angular correlation function
for the 4400 candidate field horizontal branch stars in a sample
covering 2300 deg$^2$ and found a distinct excess of pairs with angular
separations less than $10'$. Similar clumpiness has been seen in the
phase-space distribution of halo stars in the form of moving groups
found in kinematic surveys (e.g Sommer-Larsen \& Christiansen, 1987;
Croswell et al, 1991; Arnold \& Gilmore, 1992; Majewski, Munn \&
Hawley, 1994).

However, it is not yet clear to what extent (if at all)
the current observations
of non-uniform distributions of halo matter reflect
the epoch of halo formation, and 
little attention has been given to procedures suitable for evaluating
the phase-space structure of the halo as a whole.
In this paper we begin to examine the nature of debris from tidal interactions
employing 
numerical simulations of the disruption of satellites by the Milky Way.
We use our simulations to determine how long substructure
from accretion events can persist in the halo. If this timescale is
significant (ie. longer than a few Gyrs) we examine the observable
properties of the debris. We test various methods for characterizing
the substructure on artificial halo distributions, 
generated from the simulations.

We describe our computational method and
explore the behavior of tidal debris in our simulations in \S 2.
We introduce the method of Great Circle Cell Counts, and apply
it to artificial halo distributions in \S 3.
Other methods for measuring structure in the halo are discussed in
\S 4.
Finally, we summarize possible implications and limitations of our results
in \S 5.

\section{The Behavior of Debris from Tidal Interactions}
\subsection{Method}
In our simulations we represent the Milky Way  by a rigid potential, 
and model each satellite with a collection of $10^4$
self-gravitating particles whose mutual interactions are calculated
using a self-consistent field code (Hernquist $\&$ Ostriker 1992).
Since the satellite mass is much smaller than that of the
Milky Way, dynamical friction and energy exchange are assumed negligible .
Interactions between the satellites will occur infrequently
so the evolution of each satellite is considered independently.

A three-component model is used for the Galaxy (Spergel, 1995), 
in which the 
disk is described by a Miyamoto-Nagai potential (1975), 
the spheroid by a Hernquist (1990)
potential and the halo by a logarithmic potential:
\begin{equation}
        \Phi_{disk}=-{GM_{disk} \over
                 \sqrt{R^{2}+(a+\sqrt{z^{2}+b^{2}})^{2}}},
\end{equation}
\begin{equation}
        \Phi_{spher}=-{GM_{spher} \over r+c},
\end{equation}
\begin{equation}
        \Phi_{halo}=v_{halo}^2 \ln (r^{2}+d^{2}).
\end{equation}
We take $M_{disk}=1.0 \times 10^{11}, M_{spher}=3.4 \times 10^{10}, 
v_{halo}= 128, a=6.5, b=0.26, c=0.7$, and
$d=12.0$, where masses are in $M_{\odot}$, velocities are in km/s
and lengths are in  kpc. This choice of parameters yields a nearly
flat rotation curve between 1 and 30  kpc and a disk scale height of
$0.2 $ kpc.
The radial dependence of the vertical epicyclic frequency
of the disk ($\kappa_z$) between 3 and 20  kpc 
is similar to that of an exponential
disk with a 4  kpc scale length. 

Initially, each satellite is represented by a Plummer (1911) model
\begin{equation}
        \Phi=-{GM \over \sqrt{r^2+r_0^2}},
\end{equation}
where $M$ is the mass of the satellite and $r_0$ is its scale length.
The central density of the model is $\rho_0=3 M/4\pi r_0^3$. 
Plummer models were chosen because
they have flat central density profiles, as do the Milky Way's satellites.

\begin{table*}
\begin{center}
\begin{tabular}{cccccc}
Model & $M$    & $r_0$      & $\rho_0$   & $\sigma$ & $\rho_0/\rho_{Gal}$ \\
      & $10^7 M_{\odot}$ & kpc & $M_{\odot}/pc^3$ & km/s & \\
\tableline
1     & 1.00E+02 & 1.84E+00 & 3.85E-02 & 4.92E+01 & 2.49E+01 \\
2     & 4.09E+00 & 8.51E-01 & 1.58E-02 & 1.46E+01 & 1.15E+01 \\
3     & 1.91E+00 & 1.11E+00 & 3.29E-03 & 8.73E+00 & 2.76E+00 \\
4     & 6.93E+00 & 6.31E-01 & 6.60E-02 & 2.21E+01 & 9.53E+00 \\
\tableline
5     & 3.27E+01  & 1.70E+00 & $\rho_2$ & 2 $\sigma_2$       & 1.15E+01\\
6     & 1.16E+01  & 1.20E+00 & $\rho_2$ & $\sqrt{2} \sigma_2$  & 1.15E+01 \\
7     & 1.44E+00  & 6.02E-01 & $\rho_2$ & $\sigma_2/\sqrt{2} $ & 1.15E+01\\
8     & 5.11E-01  & 4.26E-01 & $\rho_2$ & $\sigma_2/2 $        & 1.15E+01\\
\tableline
9     & 5.78E+00  & 1.21E+00 & $\rho_2$/2  & $\sigma_2$  & 5.77E+00\\
10    & 3.34E+00  & 6.95E-01 & 3$\rho_2$/2 & $\sigma_2$  & 1.73E+01\\
11    & 2.89E+00  & 6.02E-01 & 2$\rho_2$   & $\sigma_2$  & 2.31E+01\\
12    & 2.36E+00  & 4.91E-01 & 3$\rho_2$   & $\sigma_2$  & 3.46E+01\\
\end{tabular}
\end{center}
\caption{Model Parameters - Column 1 labels the models; Columns 2-4 give the
mass ($M$), scale length ($r_0$), and central density ($\rho_0$) 
of the Plummer model used for the
initial distribution of particles in each satellite; Column 5 is the
characteristic velocity dispersion ($\sigma_c$)
of the satellite (see text); Column 6
gives the ratio of the satellite's initial central density to the mean
density of the Galaxy  ($\rho_{Gal}$) within the pericenter of its orbit.
For Models 5-12, $\rho_2$ and $\sigma_2$ are the
$\rho_0$ and $\sigma_c$ of Model 2 respectively. }
\end{table*}

We ran twelve different simulations whose properties are 
summarized in Table 1. 
Columns 2-4 of Table 1 give the mass, scale length and central density 
of the Plummer models.
Column 5 gives a characteristic velocity dispersion, $\sigma_c$,
for each satellite which is calculated from $\sigma_c=\sqrt{GM/r_0}$.
This quantity is interesting because we expect $\sigma_c$
to affect the rate at which tidal debris from each
satellite disperses (see \S 2.4).
The final column in Table 1 compares the 
satellite's central density to the average
density of the Milky Way 
within the pericenter of its orbit ($\rho_{Gal}$).
Analytic investigations of tidal encounters indicate that
the disruption of a satellite depends on its density (e.g.
King's tidal radius formula [King 1962]), and imply that 
the larger the value of $\rho_0/\rho_{Gal}$ for a satellite on a given
orbit, the longer it is likely to survive (see \S 2.2).

\begin{figure}
\plotfiddle{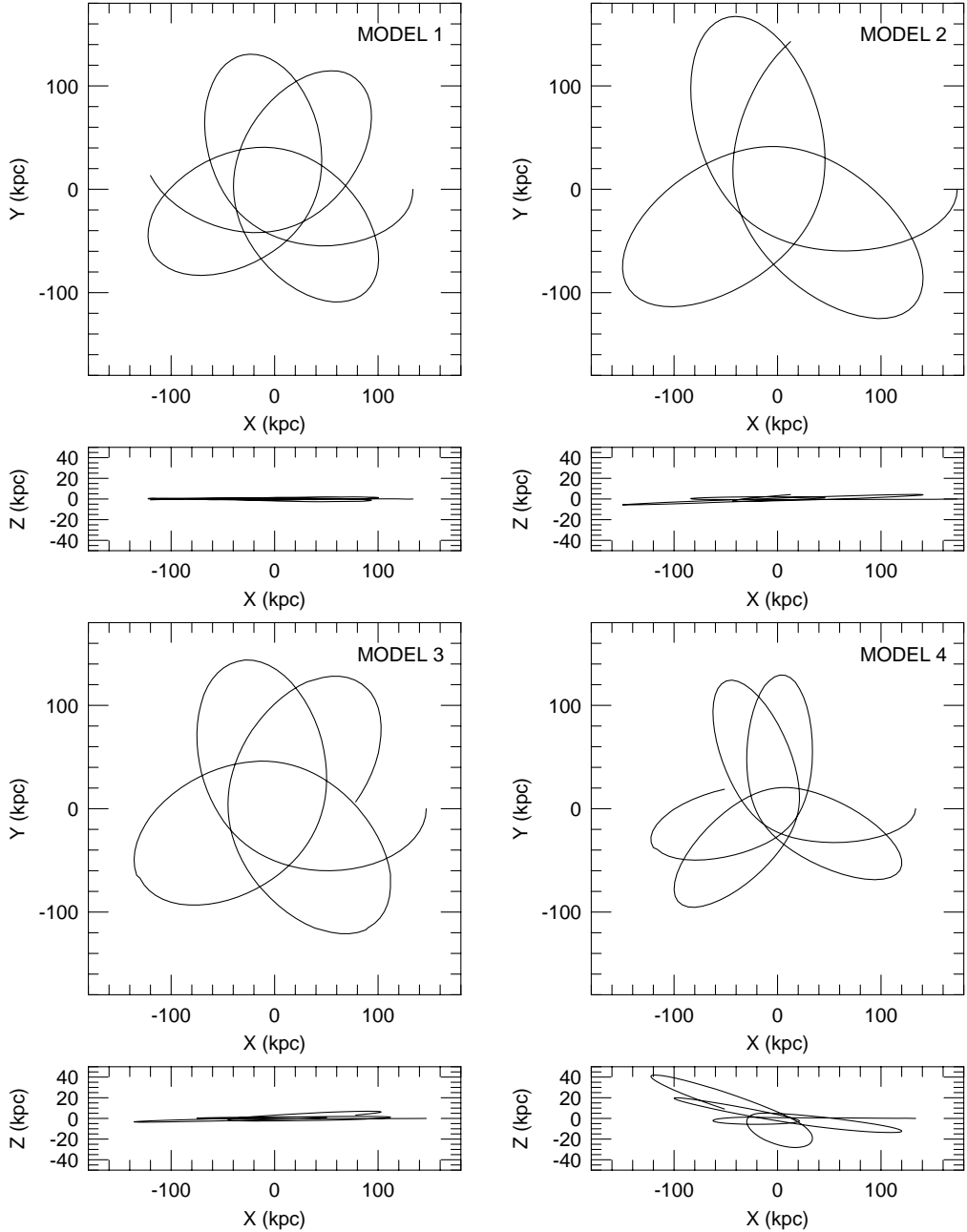}{8.in}{0}{80}{80}{-252}{0}
\caption{Orbits for Models 1-4. X and Y are co-ordinates projected
onto the plane defined by the initial position and velocity of the
satellite. Z is perpendicular to this plane.}
\end{figure}

The first four models (1-4) have satellite and orbit
parameters chosen at random to explore a range of possible outcomes. 
The orbits for each model are plotted in Figure 1. In each
case the $X-Y$ plane is defined by the satellite's initial radius
and velocity vectors and $Z$ is the position perpendicular to this plane. 
In no case is the $X-Y$ plane coincident with the Galactic disk since
the orbital parameters were chosen at random.

The remaining eight models were chosen to isolate
dynamical effects arising from the first four cases, using
Model 2 for comparisons.
These models all employ the same orbit as Model 2.
In Models 5-8  the satellites 
have the same $\rho_0$ as Model 2, 
but different $\sigma_c$.
In Models 9-12 the satellites 
have the same $\sigma_c$ as Model 2, 
but different $\rho_0$.

All computations were performed on the T3D at the Pittsburgh Supercomputing
Center. 
The parallel structure of the T3D was used to run several simulations 
at once, one on each node. 

\subsection{Tidal Disruption.}
\begin{figure}
\plotfiddle{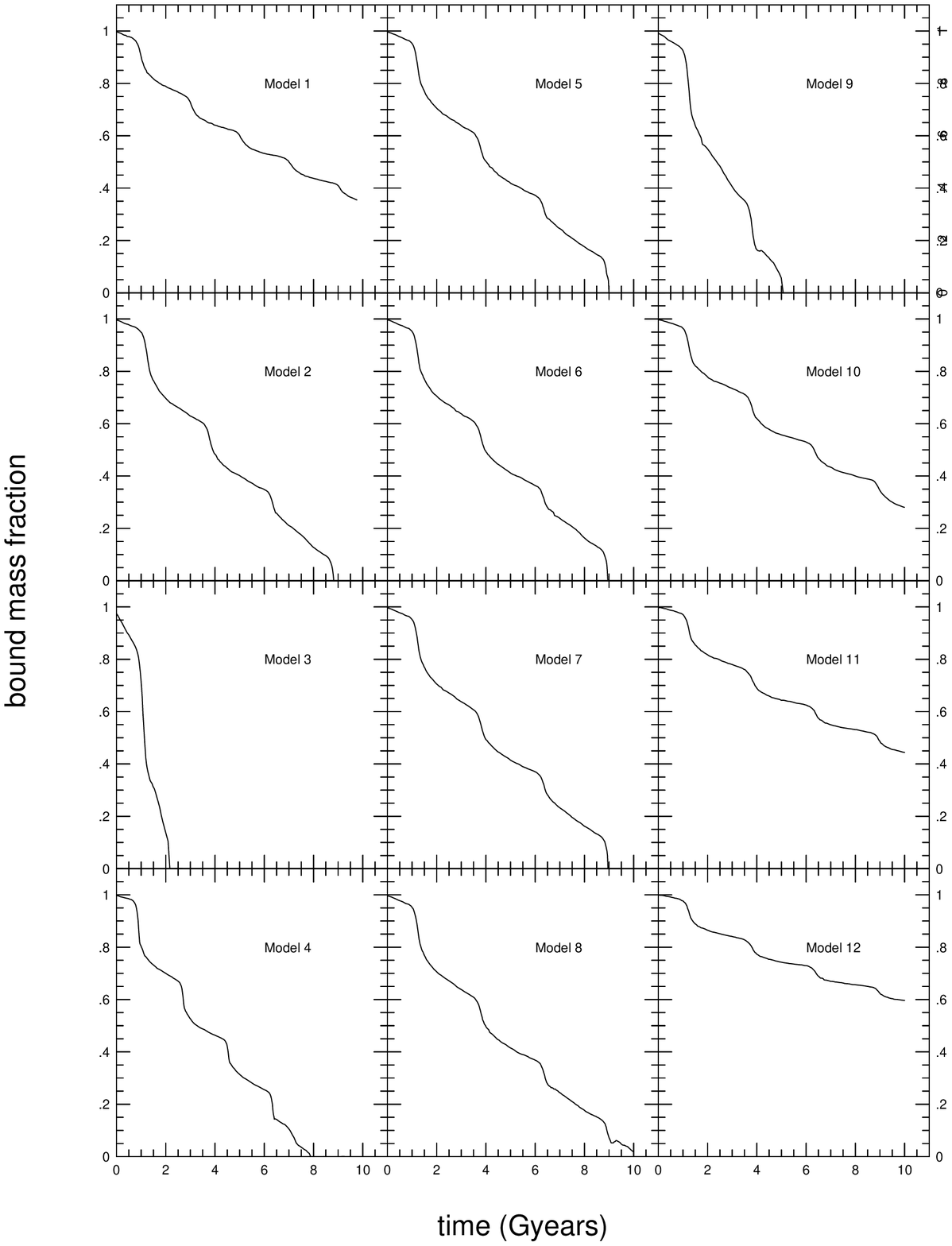}{8.in}{0}{80}{80}{-252}{0}
\caption{Fractional mass bound to each satellite as a function of time for
Models 1-12}
\end{figure}

Figure 2 shows the bound mass fraction as a function of time for
all twelve models. Comparing the time it takes for each model to
disrupt ($T_{dis}$) with the last column ($\rho_0/\rho_{Gal}$) of Table 1,
we see that $T_{dis}$ increases monotonically with 
$\rho_0/\rho_{Gal}$. This confirms the use of the density
contrast as a general guide to the fragility of a satellite in
a given orbit. Models 2 and 5-8 provide a simple demonstration
of this trend. These models all employed the same orbit and 
$\rho_0$ (and hence $\rho_0/\rho_{Gal}$) while $\sigma_c$ was varied.
In these models the satellite's masses 
span a range of nearly two orders of magnitude,
yet the mass loss rates for all five are virtually identical.

\clearpage
\subsection{Moving Groups as a Consequence of Disruption}
\begin{figure}
\plotfiddle{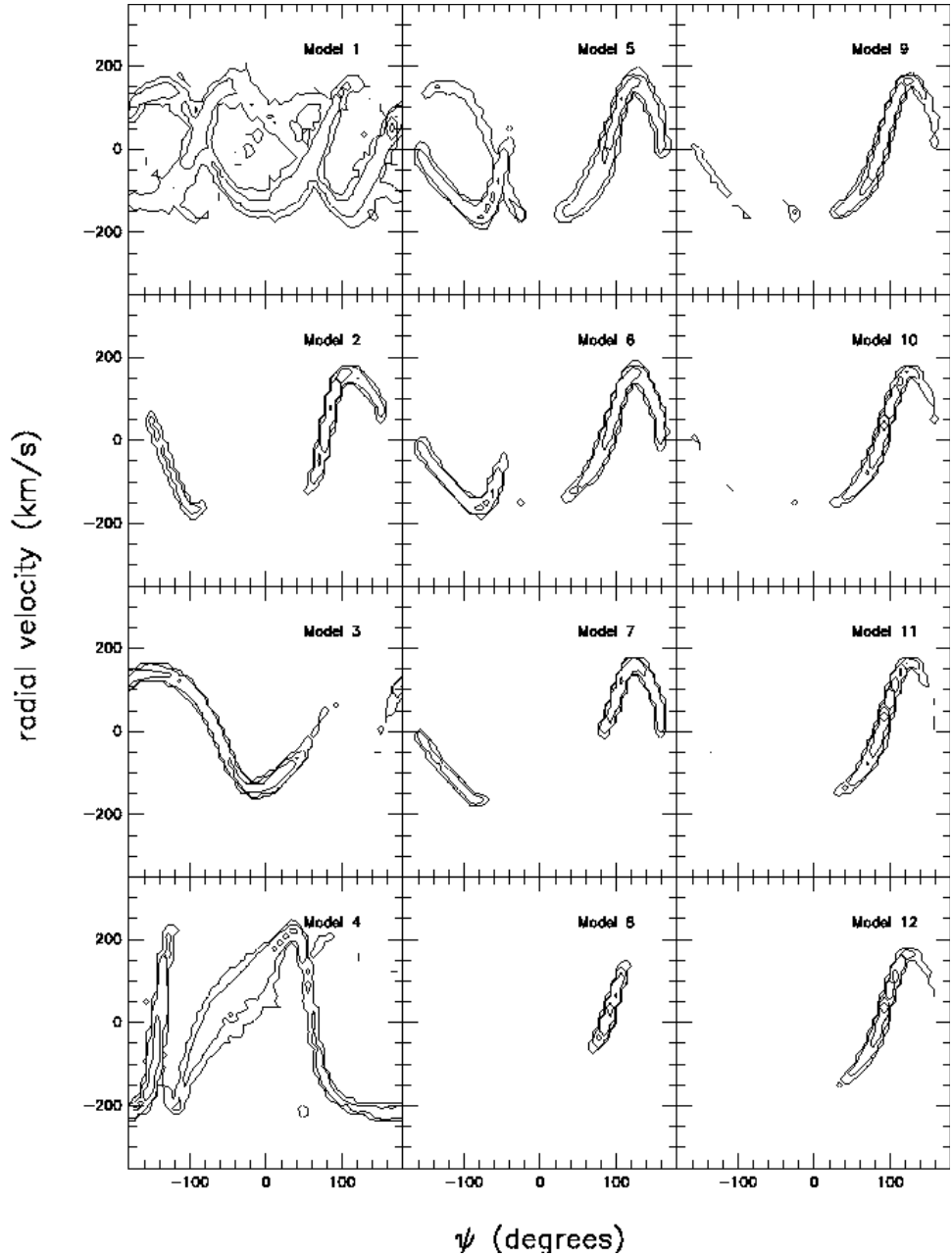}{8.in}{0}{80}{80}{-252}{0}
\caption{Contours of particles density in $ \psi - v_r$ space, where $v_r$ is
the radial velocity with respect to the center of the Galaxy and $\psi$ 
is the angle along the great circle defined by the satellite's initial 
position and velocity.}
\end{figure}

To characterize how tidal debris disperses in phase-space
after it has been stripped
from a satellite we examine the particle distribution of each
model after 10 Gyrs, from a Galactocentric viewpoint, 
in the $(\Psi,v_r)$-plane, where $\Psi$ is an
angle measured along the great circle defined by the satellite's
initial position and velocity, and $v_r$ is radial velocity.
Along an orbit, $v_r$ is an oscillatory function of $\Psi$, with zeros
at pericenter and apocenter, and an amplitude determined by 
eccentricity.
In a general spherical potential, the angular period of this radial 
oscillation is less than $2\pi$ 
and an orbit followed beyond a single radial 
oscillation is represented in the $(\Psi,v_r)$-plane by identical curves 
offset in phase.

Figure 3 shows  contours of the particle density in the $(\Psi,v_r)$
plane for Models 1-12 after 10 Gyrs. 
In all cases, the particles remain in narrow streams over
the lifetime of the Galaxy and the orbits of the satellites 
can clearly be traced, as described in the previous paragraph. 
Identical curves offset in phase are seen in Models 1 
and 5, where the tidal debris has dispersed 
beyond a single radial oscillation of the orbit.

Figure 3 offers a qualitative assessment of the nature of moving groups
associated with each model. Each value of $\Psi$ along a great circle 
gives the direction of a single line of sight, and the width of the contours
in radial velocity $\Delta v_r$ is a measure of the spread in velocities
of the debris along that line of sight. This can vary substantially along
a great circle depending on the phase of the orbit sampled by the line
of sight, suggesting that moving groups of stars 
that are related only loosely 
in distance and radial velocity 
may still be relics from a single minor accretion 
event. A systematic trend in velocity with distance should be
observed in this case.

\subsection{Debris Dispersal and the Persistence of Alignments.}
\begin{figure}
\plotone{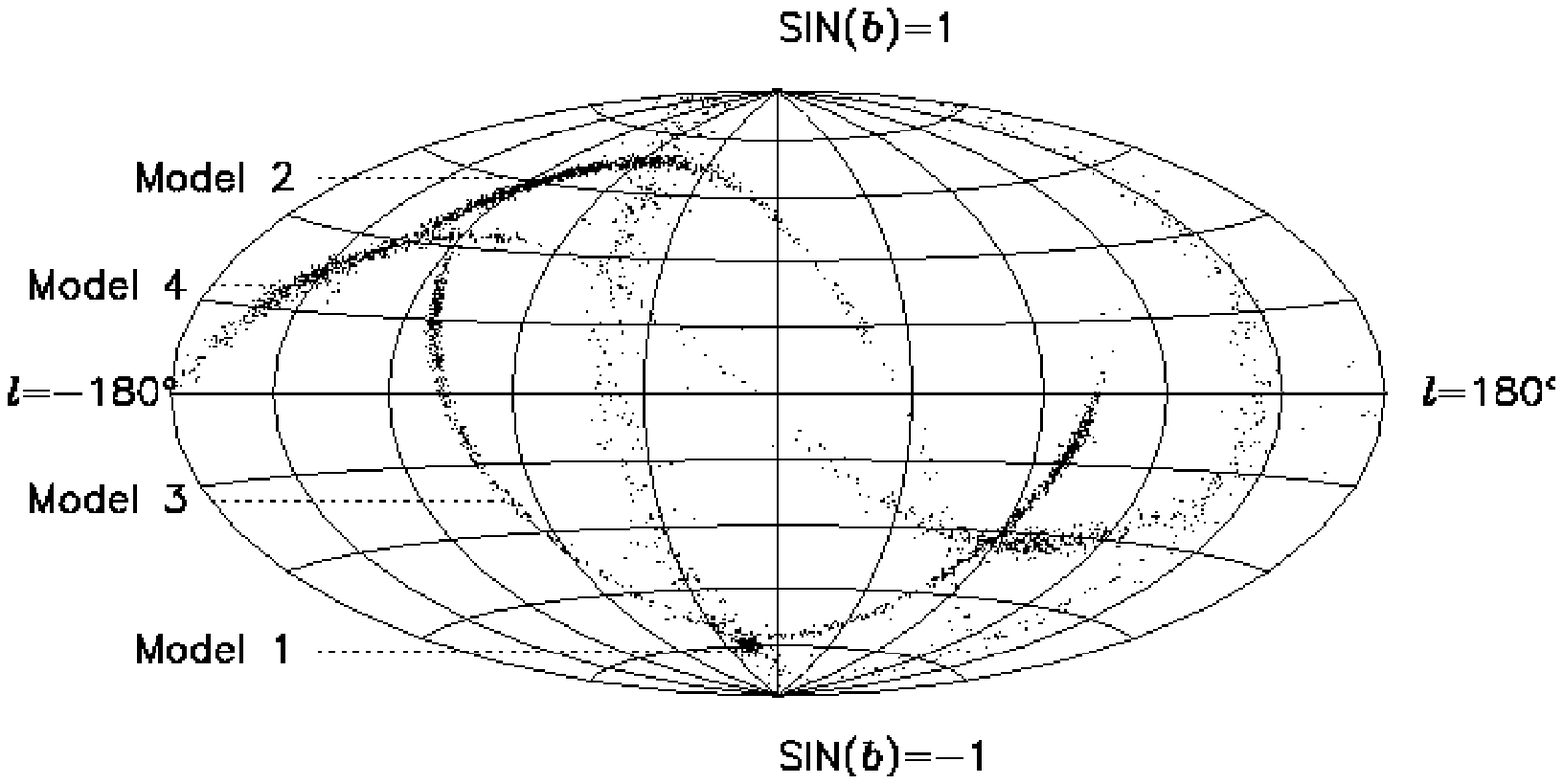}
\caption{Hammer-Aitoff full sky projection of the positions of particles
from Models 1--4  after 10 Gyrs,
in Galactocentric longitude $l$ and latitude $b$.}
\end{figure}
Figure 4 shows the positions of particles in Models 1--4
after 10 Gyrs, on a Hammer-Aitoff full sky projection, where ($l,b$) are
Galactic longitude and latitude from a Galactocentric viewpoint.
Note that the debris is aligned along tidal streamers in all cases.
In particular, the streamers from Model 1, which is nearly on an
exactly polar orbit, can be seen to
lie along a single great circle outlined by the
lines of constant longitude. 

\begin{figure}
\plotfiddle{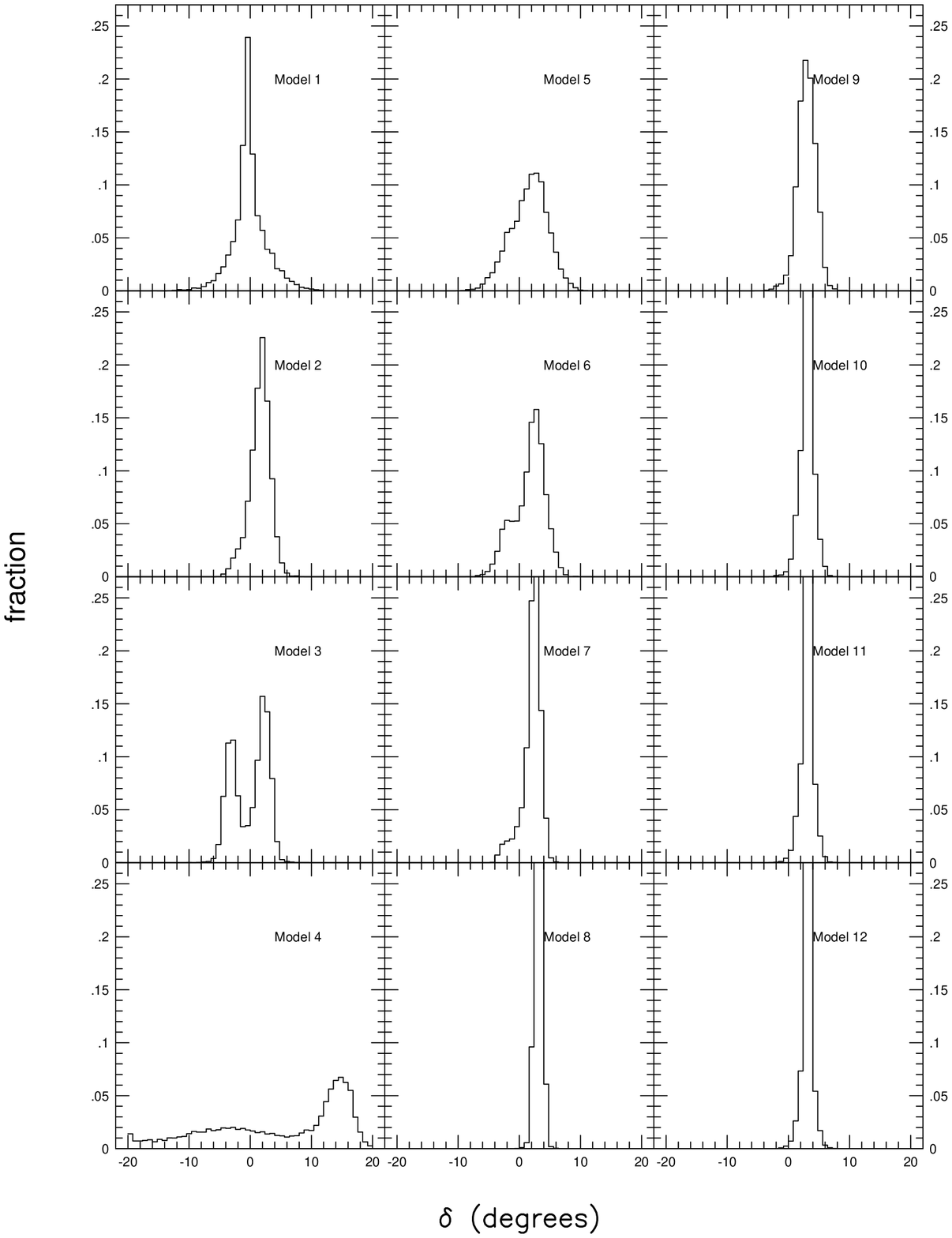}{8.in}{0}{80}{80}{-252}{0}
\caption{Number fraction of particles at an angle $\delta$ from the
great circle defined by the satellites initial position and velocity
in bins of $\Delta\delta = 1\deg$.}
\end{figure}

Figure 5 shows the fraction of particles at an angular distance 
$\delta$, perpendicular to each model's great circle, after 10 Gyrs.
The persistence of alignments of
debris along great circles can be assessed by looking at the 
{\it width} of this distribution in $\delta$, while the rate of dispersal
of debris along the orbit can be seen in the {\it length} of the
tidal streamers in $\Psi$ in Figure 3. 
Comparing Figures 3 and 5, we find that 
the extent of the tidal streamers in $\Psi$ is very much greater than
in $\delta$, despite the fact that the velocity dispersion in each satellite 
is isotropic and the debris may (naively) be expected to disperse over
similar distances in any direction.
However, the extent of the debris in both
$\delta$ and $\Psi$ increases with $\sigma_c$ of the parent satellite:
Models 5-8 have identical fractional mass loss rates and 
orbits, yet the debris covers a progressively smaller range in 
$\delta$ and $\Psi$ as $\sigma_c$ decreases;  
Models 9-12 have different fractional mass loss rates along identical 
orbits, yet the debris covers a similar range in $\delta$ and $\Psi$ 
since $\sigma_c$ is constant for these models.

\begin{figure}
\plotfiddle{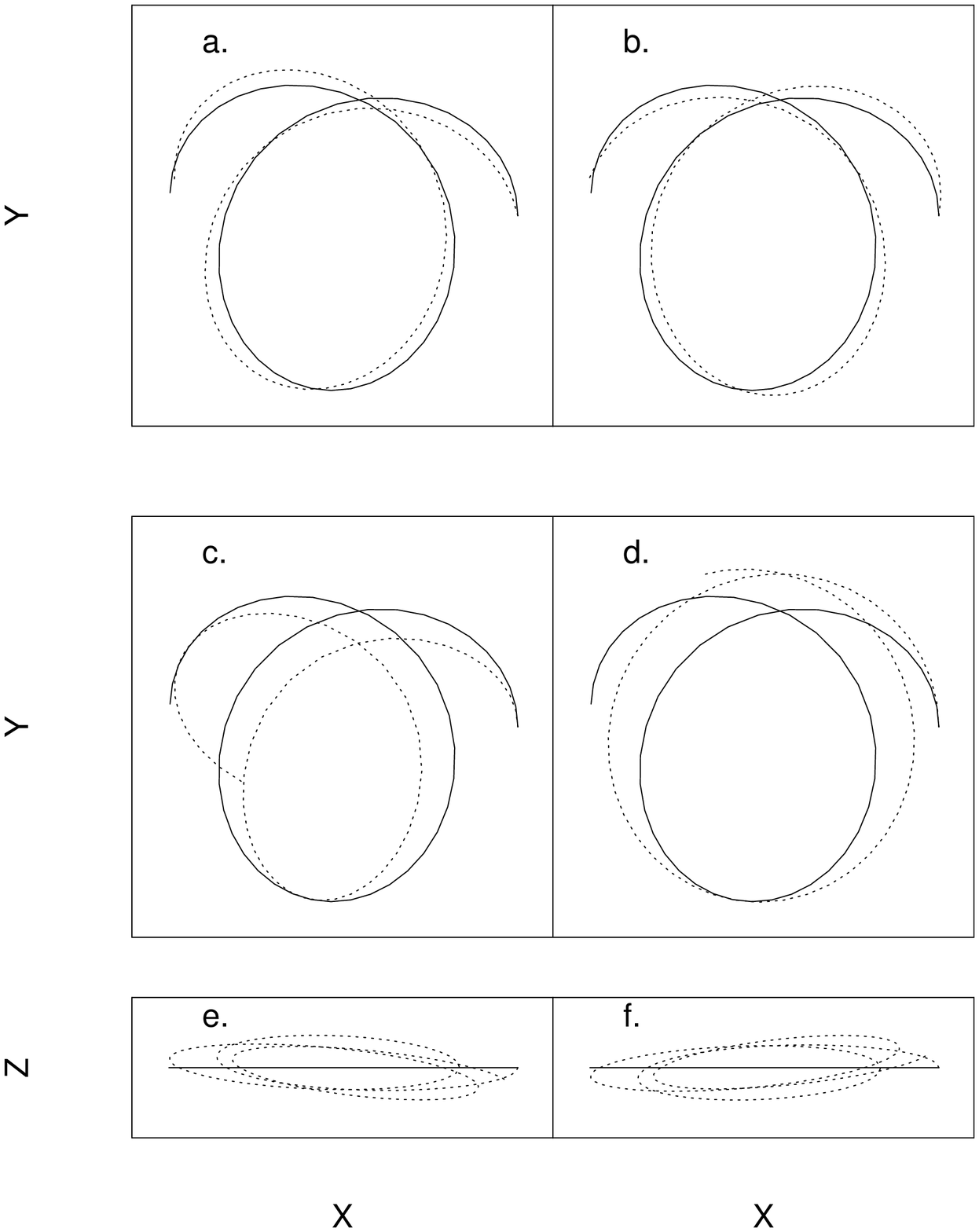}{8.in}{0}{80}{80}{-252}{0}
\caption{A rosette orbit in the potential given by equations 1--3
(solid line) compared to one perturbed by $\Delta v= \pm 0.2 v$
(dotted line) in the $R$ (a. and b.), $\phi$ (c. and d.) and
$z$ (e. and f.) directions, 
where $v$ is the initial velocity of the undisturbed orbit.}
\end{figure}

These characteristics of debris dispersal can be 
explained physically by analogy to the orbits of test particles
in a logarithmic potential which has a constant circular velocity,
$v_{c}$, over all radii. For simplicity, consider a particle 
initially on a circular orbit at radius $R$ and angular velocity
$\Omega=v_{c}/R$, and define polar coordinates
$(R,\phi)$ to lie in the orbit plane and $z$ perpendicular to it.
If the velocity of this particle is
perturbed by an amount $\Delta v =\alpha v_c$ (where $|\alpha| << 1$)
its subsequent motion can be represented using epicycles,
with the particle 
performing independent simple harmonic oscillations 
in the $R, \phi$ and $z$ directions about a guiding center on a 
circular orbit (e.g. Binney \& Tremaine 1987).
The restoring force for these oscillations is provided
by the effective potential in this frame, which arises from
a combination of the gravitational field and 
the centrifugal force due to the rotating coordinate system.
If the velocity perturbation is only in the $z$-direction, 
its new orbit can be described by an oscillation, of z-amplitude $\alpha R$,
about a guiding center following the undisturbed circular orbit.
If instead its velocity is perturbed in the $\phi$-direction, the
new orbit can be described by an oscillation, of radial amplitude 
$({\alpha/2 \over 1-\alpha/2})R$ 
around a new guiding center on a circular orbit at radius 
$({1 \over 1-\alpha/2})R$,
which moves with angular velocity $(1-\alpha/2)\Omega$. 
Hence, if a satellite with
characteristic dispersion $\sigma_c$ were disrupted along the initial
circular orbit, after $n$ orbits we may expect the debris to 
extend over an angular
distance $\alpha \propto \sigma_c/v_c$ 
in $\delta$ and $2\pi\alpha n$ in $\Psi$ .
A similar argument can explain the radial extent of the debris.
Of course, the orbits used in our simulations are far from circular.
Figure 6 shows a 
planar test particle orbit in the potential given by equations 1-3
(solid line in panels a-e). The dotted lines show orbits disturbed
by $\Delta v = \pm 0.2 v$ (where $v$ is the initial velocity of the
unperturbed orbit) in the $R$ (panels a and b), $\phi$
(panels c and d)
and $z$ (panels e and f) directions.
The qualitative behavior is identical 
to that expected for perturbations from a
circular orbit, with the particle oscillating about a guiding
center on the original orbit for $R$ and $z$ perturbations, but evolving to
a different orbit with a distinct azimuthal period for perturbations
in the $\phi$-direction.

The one exception to this picture 
is Model 4, whose debris spreads over a wide range in $\delta$
despite its relatively low $\sigma_c$.
A simple explanation is provided by Figure 1, which shows that Model
4's orbit is far from planar, since its pericenter is smaller than
the other models and its orbit precesses due to  the non-spherical potential
of the disk. Hence, although Figure 3 suggests that debris from this
interaction remains well correlated in streamers, the orbit
itself oscillates around the original great circle plane, spreading
the debris to large $\delta$, as  seen in Figure 5.

We conclude that debris trails can remain aligned in streamers near 
the parent satellite's original orbit over the lifetime of the Galaxy 
if $\sigma_c$ is not comparable to the orbital velocity.
If the orbit itself is near planar, the trails may coincide with
great circles in the sky.

\section{Great Circle Cell Counts (GC3): 
A Method for Detecting Debris Trails}

\subsection{Method}

\begin{figure}
\plotfiddle{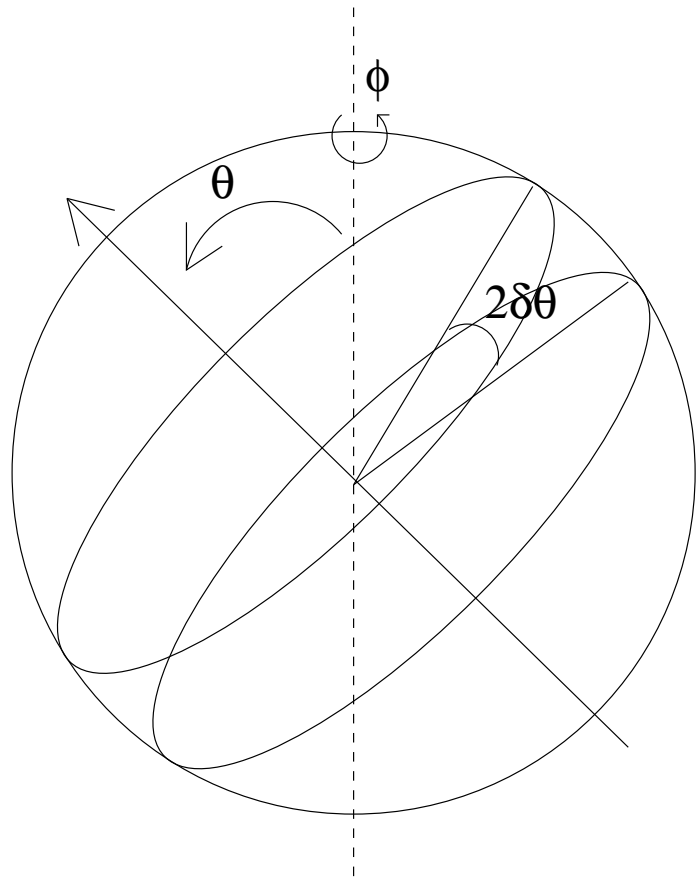}{5.in}{0}{100}{100}{-144}{0}
\caption{A great circle cell, defined by the position of its pole relative
to the symmetry axis of the Galaxy}
\end{figure}

In a spherical halo model, we find that
debris trails from the disruption of a satellite
can remain aligned along the great circle associated with the satellite's
orbit for many Gyrs (see \S 2). 
The method of Great Circle Cell Counts (hereafter GC3) 
is designed to search for this distinctive structure in a survey
of halo stars, where the stars associated with a debris
trail may contribute a negligible fraction of the total surface density.

A great circle cell is defined by the direction of its pole relative
to the North Galactic Pole, given by the two angles $(\theta,\phi)$, and
a width, $2 \times \delta \theta$, as illustrated in Figure 7.
A grid of great circle cells is chosen whose poles are equally spaced
in the intervals $-\pi<\phi<\pi$ and $0<\cos(\theta)<1$, 
to provide a systematic search for debris trails along all possible 
great circles.
The fraction of sky covered by any one cell is $p=\delta \theta$.
Hence, if $N$ stars are distributed at random on the sky, the 
number of stars to fall in a cell follows a binomial distribution
$(N,p)$, with a predictable average $\bar N=Np$ and dispersion $\sigma_{ran}=
\sqrt{Np(1-p)}$.
If some fraction of the $N$ stars are distributed preferentially
along a particular great circle, with pole $(\theta_0,\phi_0)$, 
the counts in the cell whose pole is closest
to $(\theta_0,\phi_0)$ will be a local maximum and
the dispersion, $\sigma$, of counts 
will be greater than that predicted for an isotropic distribution,
$\sigma_{ran}$.

\subsection{Results}

\begin{figure}
\plotfiddle{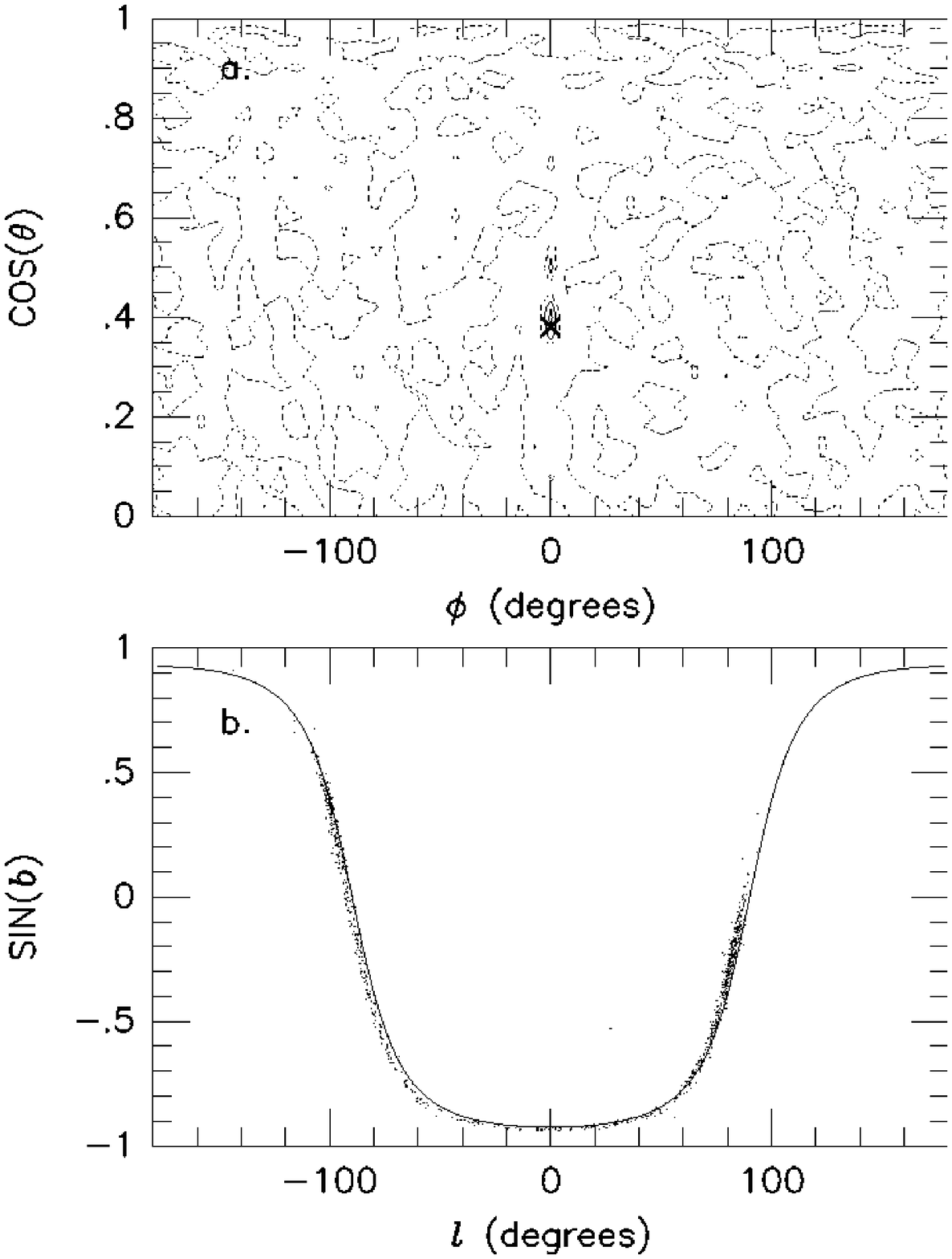}{7.in}{0}{70}{70}{-234}{0}
\caption{a. Contours of GC3 (equation 5) on a $51\times51$
grid of poles in ($\cos(\theta),\phi$) (see Figure 7) 
at levels 0 (dotted), 4, 5
and 6 (solid) for the Galactocentric view of an artificial halo 
distribution containing 1\% of its particles in a single debris trail.
The cross marks the maximum in the cell counts.
b. The positions of particles in the trail in Galactic longitude
$l$ and latitude $b$, compared to the great circle recovered from a.}
\end{figure}
We tested GC3 on artificial halo distributions containing $N_{ran}$
points distributed isotropically on the sky and $N_{sat}$ debris trails.
Each trail was based on a set of 200 points taken from the final
distribution of particles in the satellite with the shortest disruption
time - and hence the longest time for debris to disperse - Model 3.
To construct $N_{sat}$ different trails, the 
set of chosen points was rotated $N_{sat}$ times,
assuming the influence of the disk on the evolution of Model 3 
to be negligible.  
For all our calculations, we used a $51\times51$ grid of great circle
cells with width $\delta \theta=0.02$ and poles equally spaced in 
$(\cos(\theta),\phi)$-space. 
The significance of the deviation of the counts $N_{count}$
in any one cell from a random distribution was quantified by 
defining 
\begin{equation}
	{\rm GC3}=(N_{count}-\bar N)/\sigma_{ran}.
\end{equation}

As a simple test of the method, we created an artificial halo containing 
a single debris trail and $N_{ran}=19800$ particles
from a random distribution and applied GC3 from a Galactocentric
viewpoint.
Figure 8a shows the contours of the cell counts at levels
GC3=0 (dotted), 4, 5 and 6 (solid).
The cross marks the maximum in GC3. 
Figure 8b demonstrates that positions of particles in Model 3
closely follow the great circle associated with this cell.  

\begin{figure}
\plotfiddle{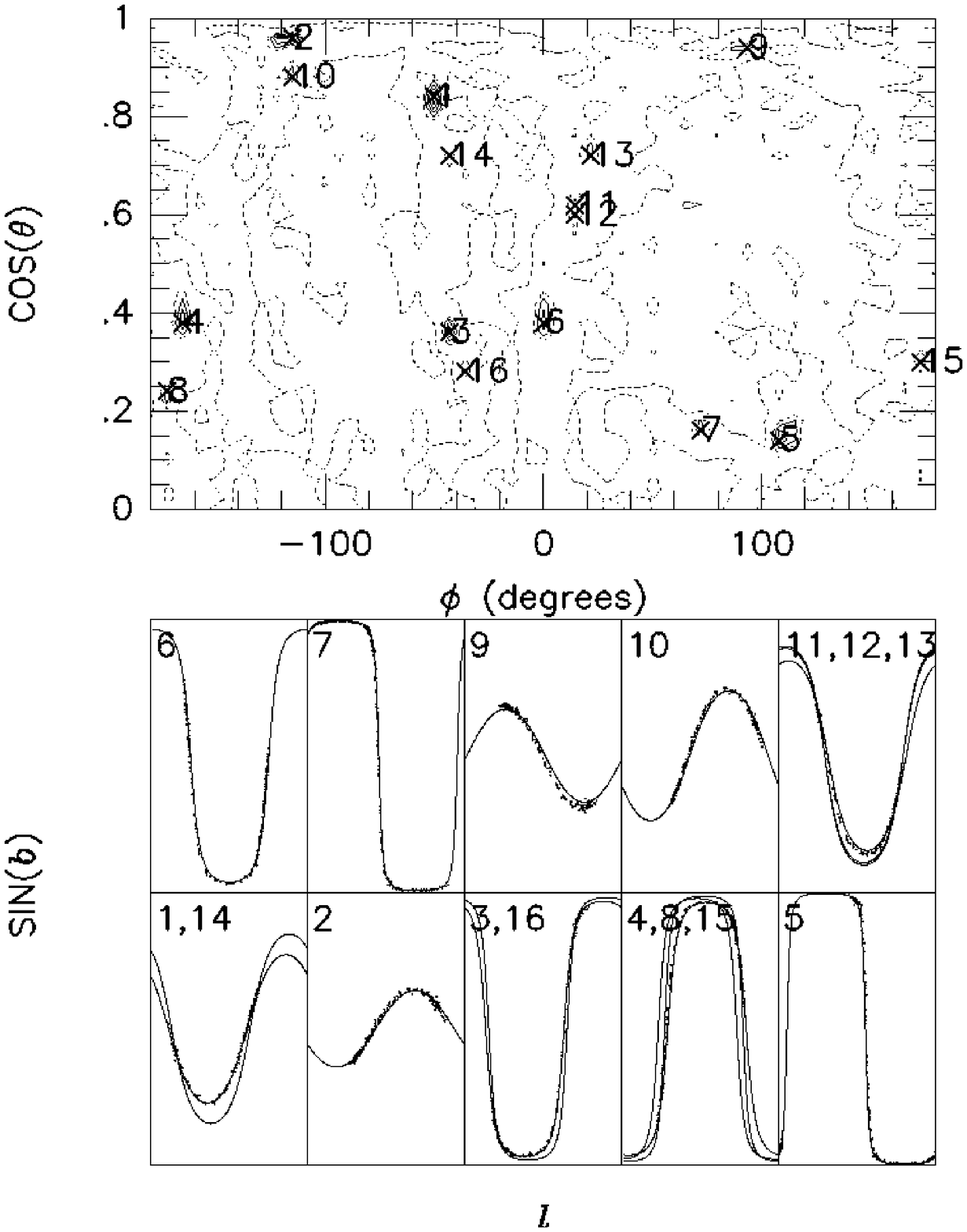}{7.in}{0}{70}{70}{-234}{0}
\caption{a. As Figure 8a, but for an artificial halo 
containing 10 debris trails, which each contribute 1\% of the particles in the
distribution. The crosses mark all local maxima above GC3 = 3 and the 
numbers give the relative importance of each maxima.
b. The positions of each debris trail compared to the most closely
associated great circles recovered from a.  }
\end{figure}
In Figure 9 we test the ability of GC3 to recover great circles
from the Galactocentric view of a halo containing $N_{sat}=10$ debris
trails, with a background of $N_{ran}=18000$ stars.
In Figure 9a the crosses mark all
local maxima above GC3=3, and the numbers label the maxima in decreasing
order of significance.
Figure 9b shows that each of the $N_{sat}$ debris trails 
can be associated with great circles recovered from the cell counts.

\begin{figure}
\plotfiddle{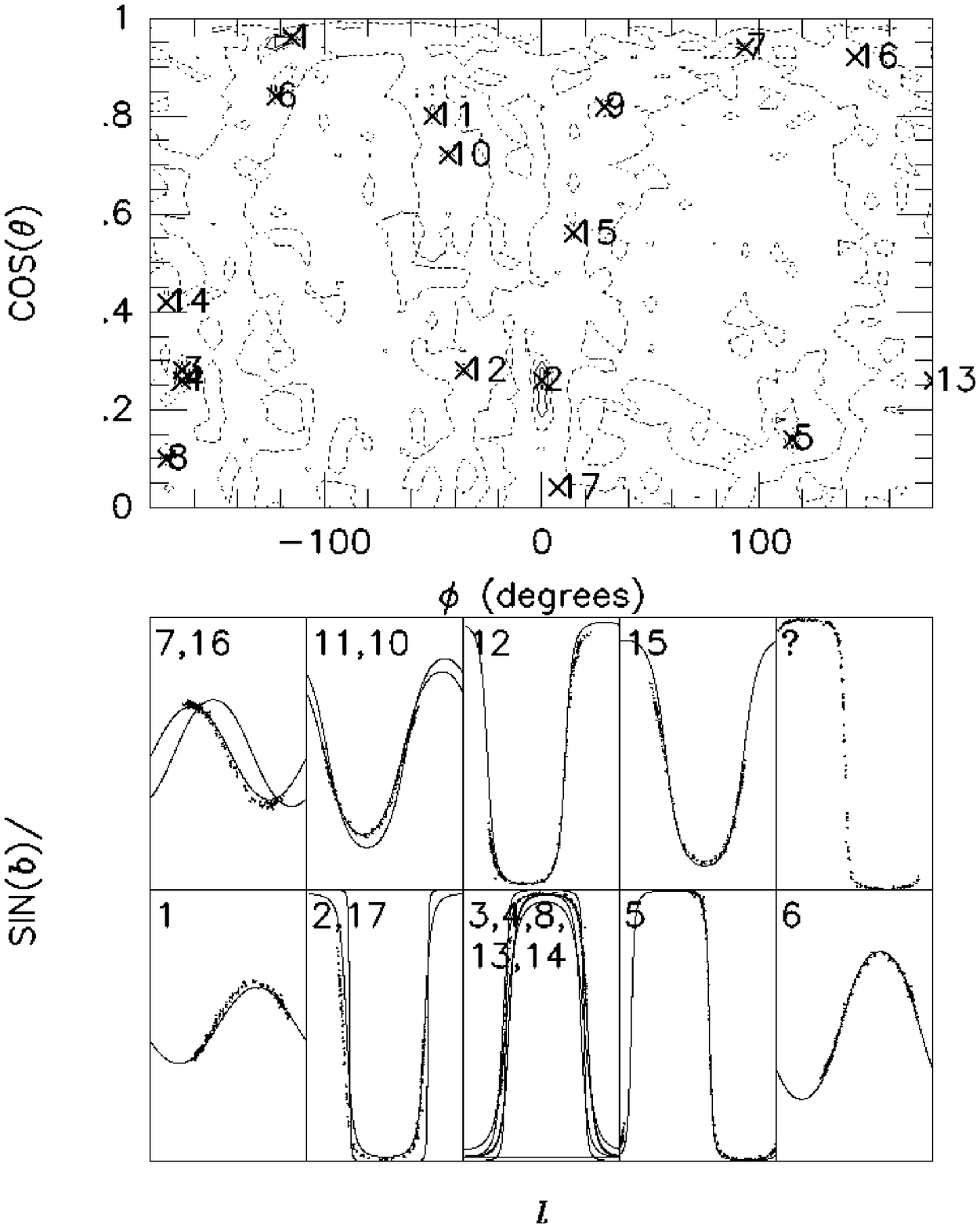}{7.in}{0}{70}{70}{-234}{0}
\caption{As Figure 9, but for the Heliocentric view of the same distribution}
\end{figure}
Unfortunately, we do not have a Galactocentric view of our own halo.
Figure 10a shows contours of GC3 from the
heliocentric view of the distribution used in Figure 9 and Figure 10b
shows that 9 out of the 10 debris trails are still successfully
recovered. 
However, the extent of the distortion of a debris trail away from 
its great circle
due to the heliocentric offset will depend on the ratio of the distance of
the Sun away from the plane of the great circle to the distance of stars
in the trail and trails at smaller distances (the pericenter of the
orbit in our case was $> 20$kpc) could be more
seriously affected.

\begin{figure}
\plotfiddle{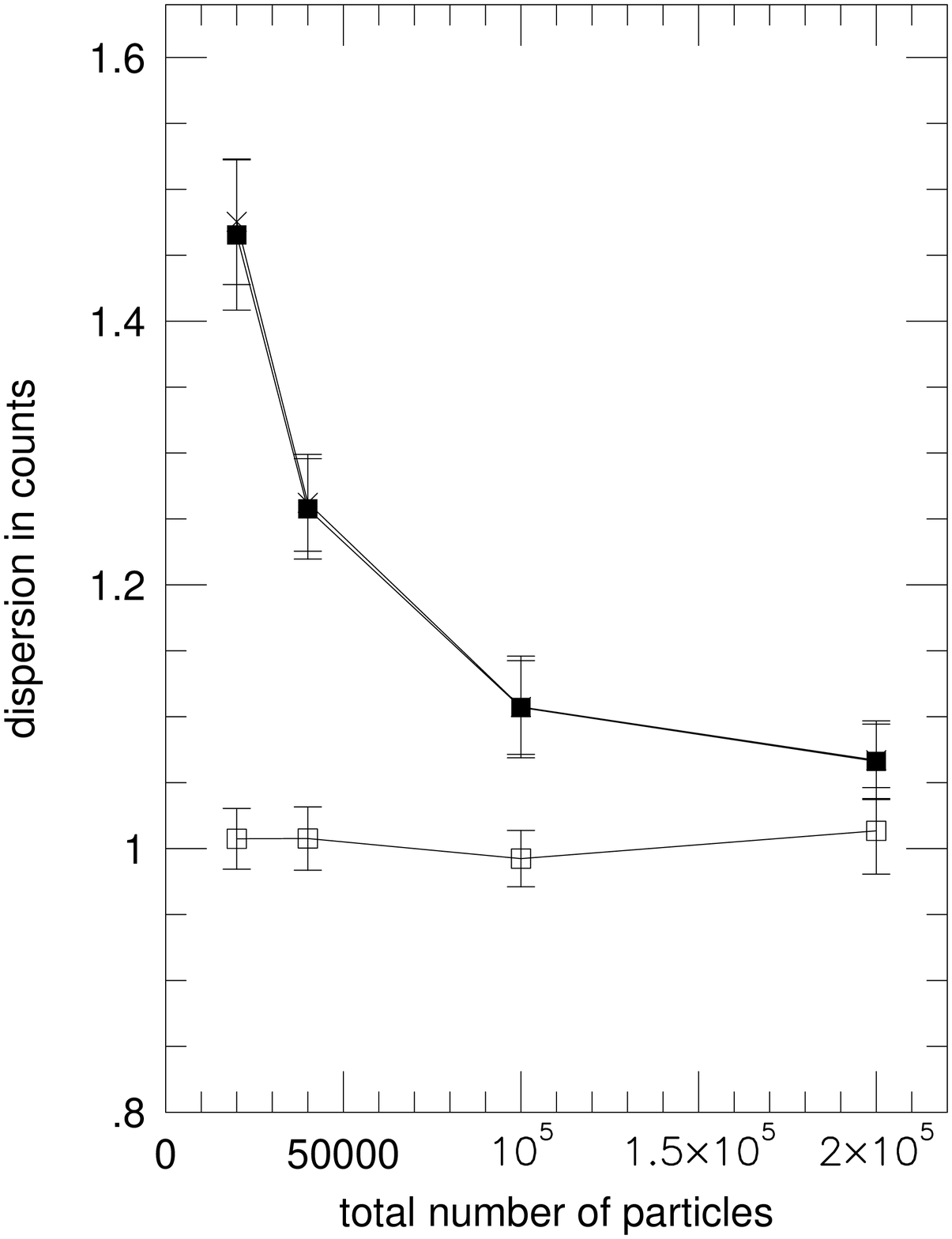}{7.in}{0}{70}{70}{-234}{0}
\caption{Dispersion in the cell counts for 
the Galactocentric (crosses) and Heliocentric (solid squares)
view of artificial halos containing the same debris trails as in
Figure 9, but with an increasing number of background stars 
($N_{tot}=N_{ran}+2000$), or decreasing percentage of accreted material.
The open squares give the values for a random distribution of $N_{tot}$
stars. The error bars give the standard deviation calculated
from 10 different realizations of artificial halos with the the same
number of debris trails and background stars.}
\end{figure}
In Figure 11 we test the sensitivity of GC3 by examining 
how the dispersion in the counts
behaves for the same distribution of accreted material used in Figures 9
and 10, but with an increasing background of randomly placed stars.
The error bars for each point 
represent the standard deviation in these quantities, calculated from 10
different realizations of an artificial halo, each containing the same number
of debris trails and isotropically distributed particles. 
The points correspond to surveys where 10\%, 5\%, 2\% and 1\% of the  stars
are relics from accretion events (or each debris trail
contains 1\%, 0.5\%, 0.2\% and 0.1\% of the stars surveyed).
From both Galactocentric (crosses) and heliocentric (solid squares)
viewpoints, the dispersion is significantly different from 
an isotropic one (open squares) in all but the last case.

In summary, the results from this section suggest that 
GC3 may be used to recover great circles associated with debris trails.
The dispersion in GC3 can
provide a statistical measure of such structure in 
a survey of halo stars.
In \S5 we discuss the application of this method to a real survey,
the effect of an oblate halo on our results and the possible inclusion of
velocity information in the method.

\section{Other Measures of Structure}

We applied other measures of structure, 
analogous to tests used in cosmological surveys
and for characterizing fluctuations in the Cosmic Microwave Background,
to artificial halo distributions with some success.

We mapped the distributions onto a grid of cells equally spaced in
($\sin(b),l$)
and found that the dispersion in the counts differed significantly
from a random distribution if as few as 1\% of the stars were taken from
debris trails. This test proved to be as sensitive as GC3 to the presence
of substructure, but had the disadvantage of not being able to distinguish
between general clumpiness and inhomogeneities specifically associated
with accretion events.

Following Doinidis \& Beers (1989), 
we calculated the angular correlation function, 
$C(\beta)$, of the distributions by counting particle pairs separated by
an angle $\beta$, and found an excess of pairs at small angles
for halos containing at least 10\% debris. This result is interesting in
that it suggests that Doinidis \& Beers' detection of power on small
scales could only result from a significant deviation from isotropy, but
again, the test fails to distinguish between clumpiness and accreted
structure.

Lastly, we calculated the Legendre Polynomial ($P_l$) decomposition of the
correlation function 
\begin{equation}
        C(\beta)=\sum_{l} {2l+1 \over 4 \pi} A_l^2 P_l(\cos(\beta)),
\end{equation}
directly from the angular distribution of the  stars
(see Peebles, 1993).
We found that for halos containing a single debris trails the 
coefficient of the even terms in the expansion
were an order of magnitude larger than the odd terms. 
Increasing the 
isotropic background increased the power only in the monopole term.
This signature results from the fact that the 
correlation function is a constant for an isotropic distribution.
and an even function for a constant density band along a great circle.
Unfortunately, the test proved impractical for
any realistic situation, since the enhancement of the even coefficients  
was lost with the introduction of more than one debris trail to the 
distribution.

\section{Discussion}
From our simulations of satellite accretion, run assuming 
a Milky Way model having a
spherical halo, we find that debris from interactions can remain aligned in
tidal streamers, close to the parent 
satellite's original orbit, for the lifetime of 
the Galaxy. 
If the orbit is near planar, the projected path of the streamers may be
closely associated with a single great circle.
The method of GC3 exploits this characteristic to recover multiple 
debris trails from artificial halos containing $\ge 1\%$ of 
material accreted from any single satellite.
The dispersion in the counts may be used to quantify the significance
of such structure in a distribution 
containing an even smaller percentage of accreted material,
and possibly as an indicator for the
role that accretion has played in the formation and evolution 
of the Milky Way.

Some of these conclusions may depend on our assumption of a spherical halo.
There is growing evidence for the oblateness of galactic halos in general
(eg Sackett \& Sparke, 1990; Sackett et al, 1994a, 1994b) and our
own in particular (e.g. Larson \& Humphreys, 1994) 
and non-planar orbits in such a potential 
may destroy the alignments with great circles seen in our simulations.
However, for orbits in moderately flattened potentials, the
total angular momentum is approximately constant and
the $z$-component of the angular momentum is exactly conserved 
(e.g. Binney \& Tremaine, 1987). 
Hence, the orbit lies near a plane whose
pole maintains a constant angle, $\theta$, with
the symmetry axis, while precessing in $\phi$ about it (see Figure 7).
This suggests that we may still
detect structure characteristic of tidal debris on our grid
of poles of great circles in ($\cos(\theta),\phi$),
either directly, or by summing over cells with constant $\theta$.
Debris from satellites on
orbits with pericenters somewhat smaller than the ones in our
models, which precess due to 
the axisymmetric potential of the disk, may also be recovered in this manner.

We base our conclusions on ``all sky surveys'' of our 
artificial halos.
In practice, it may be difficult to use
GC3 at low galactic latitudes ($|b| < 15\deg$) because 
of the predominance of disk stars
and inhomogeneity of absorption by the interstellar medium
near the Galactic plane.
However, GC3 could equally well be applied to
a restricted survey, since
the area of intersection of any great circle cell with a
given region (or regions) of sky can always be calculated.
The average and dispersion in the number of stars to fall in that
area from a random distribution of stars in the region
can be predicted, and the significance of the counts (from
equation 5) still assessed.
A survey
may have to cover a minimum of a few hundred square degrees to detect
local structure since we expect tidal debris to span several degrees,
while a survey covering a significant fraction of the sky would be required
to look at the global structure of the halo.

Finally,
our chances of detecting debris could only be improved with the addition
of velocity or distance information.
Proper motions could be used to reduce the noise in the cell counts,
by applying a stricter membership criterion for a cell, 
defined not only by a star's position, but also by the alignment of 
its velocity vector along the great circle.
Alternatively, radial positions and velocities could 
be used to refine the cell counts, using our knowledge
that these quantities should oscillate in $\Psi$ along a great circle,
with a period $P < 2\pi$.
The signature of a debris trail in a great circle cell could be
detected by systematically searching possible periods
for excess power in the quantity $A^2+B^2$ where
$$
	A=\sum y \, \cos(\Psi/P), \,\,\, B=\sum y \, \sin(\Psi/P),
$$
the sum is performed over all particles in the cell and $y$ is either
radial position or velocity.
In all three cases, once an overdense cell has been identified, the extra
coordinate (proper motion, radial position or velocity) could be used
to recover
the parent satellite's content and orbit 
by identifying coherent streams of stars along this great circle.

To summarize: we propose that the method of GC3 might be used to detect
signatures of ancient accretion events in either a spherical or oblate
halos, from surveys covering more than a few hundred square degrees.
The ultimate test of this assertion is the application of GC3
to a real survey of halo stars, such as the APM survey
(eg Maddox et al 1990), 
the APS survey of POSS-I (eg Pennington et al 1993)
or the digitized survey of the POSS-II catalogue (eg Weir, Fayyad \& 
Djorgovski 1995).

\acknowledgments We would like to thank David Spergel for helpful
discussions, and John Dubinski for an introduction to the T3D.
This work was supported in part by the Pittsburgh
Supercomputing Center, the Alfred P. Sloan Foundation, NASA
Theory Grant NAGW-2422, NSF Grants 90-18526, ASC 93-18185 and AST
91-17388, and the
Presidential Faculty Fellows program.

\end{document}